\documentclass[sn-mathphys,Numbered,iicol]{sn-jnl}

\usepackage{graphicx}%
\usepackage{multirow}%
\usepackage{amsmath,amssymb,amsfonts}%
\usepackage{amsthm}%
\usepackage{mathrsfs}%
\usepackage[title]{appendix}%
\usepackage{xcolor}%
\usepackage{textcomp}%
\usepackage{manyfoot}%
\usepackage{booktabs}%
\usepackage{algorithm}%
\usepackage{algorithmicx}%
\usepackage{algpseudocode}%
\usepackage{listings}%

\theoremstyle{thmstyleone}%
\theoremstyle{thmstyletwo}%

\theoremstyle{thmstylethree}%

\begin{document}

\title[Underestimating the uncertainty of aggregated results: the case of $W$-Boson mass]{Underestimating the uncertainty of aggregated results: the case of $W$-Boson mass}


\author*[1]{\fnm{Fintan} \sur{Costello}}\email{fintan.costello@ucd.ie}

\author[2]{\fnm{Paul} \sur{Watts}}\email{paul.watts@mu.ie}
\equalcont{These authors contributed equally to this work.  \\ }

\affil*[1]{\orgdiv{School of Computer Science}, \orgname{University College Dublin}, \orgaddress{\street{Belfield}, \city{Dublin 4}, \postcode{D04 V1W8}, \country{Ireland}}}

\affil[2]{\orgdiv{Department of Theoretical Physics}, \orgname{National University of
		Ireland  Maynooth}, \orgaddress{ \city{Maynooth}, \postcode{W23 F2K8}, \state{Co. Kildare}, \country{Ireland}}}


\abstract{Estimates of uncertainty or variance in experimental means are central to physics.  This is especially the case for `world averages' of fundamental parameters  in particle physics, which aggregate results from a number of experiments to express  current  knowledge about these parameters and where variances in these world averages reflect  uncertainty in that knowledge. The standard aggregation method used in Particle Data Group reports to estimate such parameters  is a form of fixed-effect meta-analysis.  One problem with the fixed-effect approach is that it assumes no random variation between experiments (that is, no variation in experimental accuracy, which becomes increasingly important as experimental precision rises).    This problem is well-known in the statistical literature, where the typical recommendation is to use random- rather than fixed-effect techniques.    We illustrate this problem by applying random-effect meta-analysis to estimates of the $W$-Boson mass. \\
	
Accepted for publication: European Physical Journal C: Particles and Fields
}

\maketitle 

\section{Introduction}\label{sec1}

In this paper we describe various problems with the statistical method used to produce  world average parameter estimates in particle physics: a  fixed-effect method used to estimate mass, charge and other parameters in Particle Data Group (PDG) reports \cite{Workman:2022ynf}.  We present, as an alternative, a random-effect method following a standard approach in the meta-analytic literature \cite{borenstein2021introduction} and adapted to the particle physics setting. 

	We illustrate these problems and our proposed alternative using the example of the $W$-Boson mass.  This example is of interest primarily because a 2022 high-precision  $W$-Boson mass estimate of $80.433 \pm 0.009$ GeV from the CDF group \cite{cdf2022high} exceeds the predicted Standard Model value of $80.357 \pm 0.006$ GeV by a statistically significant amount (a $6.8 \sigma$ difference) and is also inconsistent with the PDG world average \cite{Workman:2022ynf} of $80.377 \pm 0.012$ GeV (a $3.7 \sigma$ difference), though a 2023 high-precision estimate from the ATLAS group showed no such inconsistencies.  We show that estimates of the $W$-Boson mass across experiments are distributed with Normal between-experiment error (as assumed in the random-effect but not the fixed-effect statistical model);
   using the random-effect approach to estimate the $W$-Boson mass, these CDF inconsistencies  are no longer significant.

\section{Two statistical models}

To derive a useful statistical estimate of some physical quantity $\mu_*$ from repeated measurements, we must consider the variation of those measurements.  We can treat this variation as arising from the effects of various `nuisance' or `noise' factors, each following some unknown distribution. 	Scientists have two complementary techniques for dealing with nuisance factors and variability in experiments.   The first is experimental control: identifying specific known nuisance factors which could have a major impact on measurements, and ensuring that the effects of these factors are minimised or taken into account.   For some nuisance factors this involves experimental design, and aims to find methods that reduce or remove the effect of such factors on estimates by, for example, shielding equipment from external influences or designing such influences away.   For other nuisance factors such reduction is not possible: in this situation experimental control focuses on estimation, and aims to assess the degree of variability associated with such nuisance factors by, for example, calibrating equipment and measurements, so that this variability can be included in experimental results.  

Experimental control aims to address known nuisance factors individually, with different methods controlling different factors.  The second technique considers the influence of nuisance factors in aggregate, and involves the statistical modelling of variance.  The most commonly applicable model assumes a large collection of unknown nuisance factors which individually have small and additive effects on measurement so that, from the Central Limit Theorem, their combined effect will follow a Normal distribution irrespective of the distributions of these individual factors.    Assuming this statistical model of Normally distributed error we can  calculate an estimate for our parameter of interest, and of uncertainty in that estimate, from observed experimental data in a way that takes these nuisance factors into account.  Note that the validity of these estimates depends on the level of experimental control; that is, on the assumption that the remaining  nuisance factors each have small effects (because major nuisance factors have been individually dealt with in some way).

How do these techniques apply to the aggregation of results across multiple experiments? In terms of experimental control, such aggregation must eliminate or adjust for any major differences between experiments (any major between-experiment nuisance factors), because such differences will have systematic effects on all measurements in a given experiment and so will produce systematic differences in experimental results.
In terms of statistical modelling of variance, two types of approach are possible.  The first assumes that, with major between-experiment differences controlled, all experiments estimate the same parameter of interest $\mu_*$ with no further differences.  In this  approach an overall estimate for this parameter can be produced by averaging experimental means weighted by the inverse of their within-experiment variance (so that the more precise estimates of $\mu_*$ contribute more to the aggregated value); in the meta-analytic literature this is referred to as `fixed-effect' estimation \cite{borenstein2010basic}.   

The second approach assumes that, even when major between-experiment nuisance factors are controlled, there remains some collection of unknown between-experiment nuisance factors which individually have small random and additive effects  (that is, where each between-experiment nuisance factor randomly shifts individual experimental means $\mu_i$, just as each within-experiment nuisance factor randomly shifts individual measurements).   From the Central Limit Theorem the combined effect of these between-experiment nuisance factors will follow a Normal distribution irrespective of the distributions of these individual factors; and so experimental means  are not all fixed at $\mu_*$ but instead are Normally distributed around $\mu_*$ with some between-experiment variance.  This is typically referred to as `random-effect' estimation, and is a standard recommendation in the meta-analytic literature  \cite{borenstein2021introduction,gurevitch2018meta}.

We describe the fixed- and random-effect approaches to estimation and aggregation of experimental results in detail below, using the following notation.   We use Greek letters (e.g. $\mu,\sigma$) to represent unknown parameter values, add `hats' to indicate estimates of these values (e.g. $\hat{\mu},\hat{\sigma}$)  and use capital Roman letters to represent experimental measurements or statistics (e.g. $X$, a measurement; $N$, the number of measurements; $S^2$, the sample variance of those measurements) with overlines used to indicate averages (e.g. $\overline{X}$ is the average of some set of measurements $X$).  When referring to (theoretical, estimated or sample) means we use a (typically numeric) subscript $i$ to refer to the mean or variance associated with experiment $i$.  We use the special `*' subscript to refer to the mean across all experiments (so $\mu_*$ is  the parameter to be estimated)  and $\sigma_*^2$ to refer to the variance of individual experimental means $\mu_i$ around $\mu_*$.

\subsection{Fixed-effect estimation}
Given $K$ experiments each with $N_i$  individual measurements $X_{i,1} \ldots X_{i,N_i}$ and so sample mean 
\[ \overline{X}_i = \frac{1}{N_i}\sum_{j=1}^{N_i} X_{i,j}\]
and  sample variance 
\[ S^2_i = \frac{1}{N_i-1}\sum_{j=1}^{N_i} \left(\overline{X}_i- X_{i,j}\right)^2\]
the statistical model in fixed-effect estimation is one where these experimental means $\overline{X}_i$ are assumed to follow the distribution
\[\overline{X}_i \sim \mathcal{N}(\mu_*,\sigma_i^2) \]
so that all  means estimate the same (fixed but unknown) parameter of interest $\mu_*$, but where each mean has a different associated variance $\sigma_i^2$.   Given this model the precision of each mean $\overline{X}_i$ as an estimate of $\mu_*$ is $1/\sigma_i^2$; and letting $\hat{\sigma}_i^2$ represent an estimate for $\sigma_i^2$ (typically the sample variance of the mean, so that $\hat{\sigma}_i^2 = S_i^2/N_i$) our estimate for $\mu_*$  is the inverse-variance weighted average of the means
\begin{equation}
\label{eq:fixed_est}
\hat{\mu}_{F} =   \hat{\sigma}^2_{F}\sum\limits_{i=1}^K \frac{ \overline{X}_i }{\hat{\sigma}_i^2} 
\end{equation} 
where the normalising value
\begin{equation}
\label{eq:fixed_var}
\hat{\sigma}^2_{F}  =\frac{1}{\sum\limits_{i=1}^K  \frac{ 1 }{\hat{\sigma}_i^2} }
\end{equation} 
is also the variance of the estimate $\hat{\mu}_{F}$.   Assuming that all experiments estimate the same mean $\mu_*$ (the fixed-effect assumption), $\hat{\mu}_{F}$ provides the most precise estimate of $\mu_*$ possible given our set of $K$ experimental means (that is, the estimate with minimum variance $\hat{\sigma}^2_{F}$).  As our number of experiments $K$ grows the variance of this estimate $\hat{\sigma}^2_{F}$ necessarily falls
because the sum
\[ \sum\limits_{i=1}^K  \frac{ 1 }{\hat{\sigma}_i^2} \]
necessarily grows with $K$ and $\hat{\sigma}_F^2$ is the inverse of this sum;  and so the precision of $\hat{\mu}_{F}$ increases with the number of experiments $K$ irrespective of the variance of the individual experiments $\hat{\sigma}^2_{i}$, and further increases as those individual variances fall.

\subsection{Random-effect estimation}  
The statistical model in random-effect estimation is one where each experimental mean $\overline{X}_i$ is assumed to follow the distribution
\[\overline{X}_i \sim \mathcal{N}(\mu_i,\sigma_i^2) \]
conditional on $\mu_i$, and where the latent means $\mu_i$ themselves follow the distribution   
\[\mu_i \sim \mathcal{N}(\mu_*,\sigma_*^2) \]
so that the unconditional or complete distribution of sample means is
\begin{equation} 
\label{eq:random_effect}
\overline{X}_i \sim \mathcal{N}(\mu_*,\sigma_i^2 + \sigma_*^2)
\end{equation}
This approach thus assumes two forms of  uncertainty or variance in sample means: variance in individual measurements around $\mu_i$, and variance in latent means $\mu_i$ around $\mu_*$.  Just as within-experiment statistical variance $\sigma^2_i$ arises from random variation or error from measurement to measurement,  between-experiment statistical variance $\sigma_*^2$ arises  from random variation or error from experiment to experiment (minor differences in equipment,  in calibration, in experimental procedure, in analysis technique, in setting and so on).   These two forms of statistical variation (within-experiment variation in individual measurements around latent means $\mu_i$, and between-experiment variation in means $\mu_i$ around the parameter of interest $\mu_*$) are connected to the ideas of precision and accuracy in experimental results: reducing within-experiment variation increases the precision of an experiment as an estimate of $\mu_i$ but not its accuracy as an estimate of $\mu_*$, while reducing between-experiment variation increases the accuracy of the experiment as an estimate of $\mu_*$, but not its precision as an estimate of $\mu_i$ \cite{mandel2012statistical}.  

Letting $\hat{\sigma}_i^2$  and  $\hat{\sigma}_*^2$ represent estimates of the respective variances, in this model the precision of each mean $\overline{X}_i$ as an estimate of $\mu_*$ is $1/(\hat{\sigma}_i^2 + \hat{\sigma}_*^2)$  and as before the  minimum-variance estimate for $\mu_*$ is the inverse-variance weighted average
\begin{equation}
\label{eq:random_est}
\hat{\mu}_R = \hat{\sigma}^2_{R} \sum\limits_{i=1}^K \frac{\overline{X}_i}{ \hat{\sigma}_i^2 + \hat{\sigma}_*^2 }
\end{equation} 
where the normalising value
\begin{equation}
\label{eq:random_var}
\hat{\sigma}^2_{R}  = \frac{1}{\sum\limits_{i=1}^K \frac{1}{ \hat{\sigma}_i^2 + \hat{\sigma}_*^2 }}
\end{equation} 
is again the variance of the estimate $\hat{\mu}_R$.

\subsection{Between-experiment variance}

The random-effect model depends, of course, on our ability to estimate between-experiment variance $\hat{\sigma}_*^2$. Just as the within-experiment variation of the sample mean $\hat{\sigma}_i^2$ can be estimated by the sample variance $S_i^2/N_i$, the between-experiment variation of means $\hat{\sigma}_*^2$ can be estimated in terms of the sample variance of the means from our set of $K$ experiments.  This estimate can be obtained in a variety of ways (see \cite{veroniki2016methods} for an extensive review); we consider the two most relevant methods here.     The first is based on Cochran's statistic  \cite{cochran1954combination} 
\begin{equation}
\label{eq:Q}
Q =  \sum\limits_{i=1}^K \frac{ ( \overline{X}_i- \hat{\mu}_F)^2 }{\hat{\sigma}_i^2}
\end{equation}
This statistic $Q$ represents the squared difference between each experiment's mean $ \overline{X}_i$ and the fixed-effect estimate $\hat{\mu}_F$, divided by that experiment's variance $\hat{\sigma}_i^2$; and so under the assumption of the fixed-effect model this statistic $Q$ follows a $\chi_{K-1}^2$ distribution at least to a rough approximation \cite{hoaglin2016misunderstandings} and so has an approximate expected value of $K-1$.  Standard tests of heterogeneity of means (that is, of deviation from the fixed-effect model) in the meta-analytic literature \cite{borenstein2010basic,borenstein2021introduction,viechtbauer2010conducting} estimate the probability of obtaining a given value of $Q$ (or a more extreme value) under the assumed fixed-effect distribution $\chi_{K-1}^2$: if this probability is low then the fixed-effect hypothesis is rejected.  A related measure
\begin{equation}
\label{eq:H}
H= \frac{Q}{K-1}
\end{equation}
is also used as a measure of heterogeneity independent of the number of experiments being considered \cite{higgins2002quantifying,higgins2009re}.

Assuming additional between-experiment variance $\sigma_*^2$, this statistic $Q$ has a moment-based expected value of approximately
\[  \left< Q \right> =  \sigma_*^2\left( \frac{1}{\hat{\sigma}^2_{F}} - \hat{\sigma}^2_{F} \sum\limits_{i=1}^K \left(\frac{1}{ \hat{\sigma}_i^2  }\right)^2\right) + (K-1)\] 
which gives the DerSimonian and Laird (DL) estimator \cite{dersimonian1986meta} for between-experiment variance $\sigma_*^2$ of
\[    \frac{Q - (K-1) }{ \frac{1}{\hat{\sigma}^2_{F}}-  \hat{\sigma}^2_{F}  \sum\limits_{i=1}^K \left(\frac{1}{ \hat{\sigma}i^2  }\right)^2}   \]
This is an unbiased estimate of $\sigma_*^2$ when the individual variance values $\sigma_i^2$ are known exactly,  but is subject to bias when they are estimated from sample data.   

An alternative approach, referred to as the Cochran estimator \cite{cochran1954combination}, the Hedges estimator \cite{hedges1983random}, or the variance-component estimator \cite{veroniki2016methods} estimates $\hat{\sigma}^2_*$ by considering the unweighted grand mean 
\[ \overline{X}_* = \frac{1}{K} \sum_{i=1}^{K}\overline{X}_{i}\]
and its sample variance
\begin{equation}
\label{eq:S}
S_*^2 = \frac{1}{K-1}\sum_{i=1}^{K}\left(\overline{X}_{i}-  \overline{X}_* \right)^2 
\end{equation}       
From our statistical model (eq \ref{eq:random_effect}) we see that the expected variance in $\overline{X}_*$ is
\begin{equation}
\label{eq:theoretical_var}
\begin{split} 
\frac{1}{K^2} \sum_{i=1}^{K}\left(\sigma^2_i + \sigma_*^2\right)= \sigma_*^2/K+ \frac{1}{K^2}\sum_{i=1}^{K} \sigma^2_i  
\end{split}
\end{equation}
and equating this expression with the sample variance of the mean $S_*^2/K$ we get an estimator for $\sigma_*^2$  of
\begin{equation}
\label{eq:sigma_star}
\hat{\sigma}^2_*  = S_*^2- \frac{1}{K}\sum_{i=1}^{K} \hat{\sigma}^2_i  
\end{equation} 
with variance or uncertainty in this estimate of
\begin{equation*} 
\begin{split}
\frac{2}{(K-1)^2}\left[ \left(1-\frac{2}{K}\right) \sum_{i=1}^K v_i^2  +\frac{1}{K^2}\left(\sum_{i=1}^{K} v_i\right)^2    \right]
\end{split}
\end{equation*}
where $v_i = \sigma_i^2+\sigma_*^2$  \cite{viechtbauer2005bias}. 

For both the DL and the variance-component estimator values $\hat{\sigma}_*^2 \leq 0$  are typically taken to indicate $\sigma_*^2 = 0$.  We don't make that assumption here as it is inconsistent with the random-effect approach (which assumes $\sigma_*^2 > 0$).  Instead  we take the random-effect assumption $\sigma_*^2 > 0$ as our default or null hypothesis, and only reject that hypothesis  when the assumed Normal distribution for means is is not supported in a statistical test of Normality such as the Shapiro-Wilk test \cite{razali2011power,shapiro1968comparative}.  In such cases we have evidence that experimental means do not all follow the same Normal distribution, and so assume that they have some hierarchical or group structure, with results from each group of experiments following a different Normal distribution (with mean and variance parameters particular to that group) and with grouping defined by commonalities between these experiments (so that in the $W$-Boson example one such group may contain all experiments that ran on a particular collider, for example).    When  there is no statistical evidence to reject the assumed Normal distribution for means, we take  values  $\hat{\sigma}_*^2 \leq 0$ to indicate that $\sigma_*^2$ cannot be estimated  because within-experiment uncertainties $\hat{\sigma}_i^2$  are too large, or the number of experimental means $K$ being aggregated is too small, to allow such an estimate.   

\subsection{Hypothesis testing }

While both random- and fixed-effect estimates involve inverse-variance weighted averages, a critical difference follows from the inclusion of a between-experiment variance estimate  $\hat{\sigma}_*^2$ in the random-effect approach.   This is connected to hypothesis testing: suppose we have some theoretical prediction about the parameter of interest, represented by the normal distribution
\[ \mu_* \sim \mathcal{N}(\mu_T, \hat{\sigma}_{T}^2)\]
and we obtain some new experimental result with sample mean  $\overline{X}_{new}$ and  sample variance of the mean   $\hat{\sigma}_{new}^2$.  

Since $\overline{X}_{new}$ is assumed to be distributed around $\mu_*$ in the fixed-effect approach, in that approach the theoretically predicted distribution for the difference $d=\overline{X}_{new} -\mu_T$ is      
\[ d\sim \mathcal{N}(0, \hat{\sigma}_{T}^2+ \hat{\sigma}_{new}^2)\]
which has a standard deviation of 
\[\sigma_d=\sqrt{\hat{\sigma}_{T}^2+ \hat{\sigma}_{new}^2}\] 
If this difference $d$ is greater than some standard criterion ($5\sigma_d$, say) we can take our observed result as inconsistent with the theoretical prediction, suggesting a  significant deviation from theory.

In the random-effect approach, however, $\overline{X}_{new}$ is assumed to be distributed around $\mu_{new}$ where $\mu_{new}$ is itself drawn from the distribution
\[ \mu_{new} \sim \mathcal{N}(\mu_*,\hat{\sigma}_*^2) \] 
and so theoretically predicted distribution for the difference $d$  is      
\begin{equation}
\label{eq:RE_test}
d \sim \mathcal{N}(0, \hat{\sigma}_{T}^2+ \hat{\sigma}_{new}^2 + \hat{\sigma}_*^2)
\end{equation} 
and this difference has a standard deviation of 
\[\sigma_{d_*}=\sqrt{\hat{\sigma}_{T}^2+ \hat{\sigma}_{new}^2 + \hat{\sigma}_*^2}\]
Since this depends on the between-experiment variance of our experimental means $\overline{X}_i$ it is larger than that obtained in the fixed-effect approach, and results which suggest a significant deviation from theory in a fixed-effect analysis may be judged consistent with theory once this between-experiment variance is taken into account.

\section{The PDG approach }  

The PDG method for estimating world averages of particle physics parameters  follows the fixed-effect approach with a number of modifications or extensions (see the PDG section on Procedures; pages 13 to 19 of \cite{Workman:2022ynf}).     First, this approach does not produce a world average for a given parameter by aggregating all experiments estimating that parameter:  instead some experiments are included in the PDG fit, while others are excluded.  Among the reasons for the exclusion of a given experiment are (quoting ref. \cite{Workman:2022ynf}, p. 14) 

\begin{quote}
	\begin{itemize}
		{\small
		\item It involves assumptions we question.
		\item It has a poor signal-to-noise ratio, low statistical significance, or is otherwise of poorer quality
		than other data available.
		\item It is clearly inconsistent with other results that appear to be more reliable. Usually we  then
		state the criterion, which sometimes is quite subjective, for selecting ``more reliable'' data for
		averaging. 
	} 
	\end{itemize}  
   
\end{quote}

Second, in the PDG approach variance in the mean for some experiment $i$ is assumed to have statistical ($\sigma_{stat_i}^2$) and systematic ($\sigma^2_{sys_i}$) components, with statistical variance estimated by the sampling variance of the mean $S^2_i/N_i$ and systematic variance estimated in various ad-hoc ways involving factors that influence the experimental mean (uncertainties in calibration, or in the $W$-Boson case uncertainties due to measurement width and radiative corrections). 

Third, given a group of $K_c$ experimental estimates from some collider $c$, the PDG approach assesses the quality of estimates coming from that collider  by calculating the aggregated fixed-effect mean $\hat{\mu}_{F_c}$ and variance $\hat{\sigma}^2_{F_c}$ for that group (Eqs \ref{eq:fixed_est} and \ref{eq:fixed_var}), taking the variance of each experiment to be $\sigma^2_{stat_i} +\sigma^2_{sys_i}$; and also calculating the $\chi^2$ sum $Q_c$ (Eq \ref{eq:Q}) and the ratio  
$H_c=Q_c/(K_c-1)$ (Eq. \ref{eq:H}).   As we saw earlier, under the fixed-effect assumption we expect $Q_c$ to have a value of approximately $K_c-1$.  If  $H_c > 1$ this expectation is violated to some degree; in the PDG approach this is taken to indicate low quality estimation for group $c$.
If $H_c$ is very large we may choose not to use this group of estimates  at all; if  $H_c$ is moderately above $1$ in the PDG approach we  may use those estimates but increase the standard deviation for each individual experiment in group $c$ by multiplying  $ \sigma_{i_c}$ by the scaling factor $H_c$.   Noting that in the PDG presentation these variables $Q$ and $H$ are referred to as $\chi^2$ and $S$ (and we use $Q$ and $H$ for consistency with the earlier presentation of the random-effect model), the reason for this rescaling is that

\begin{quote} `the large value of the $\chi^2$ is likely to be due to underestimation of errors in at least one of the experiments. Not knowing which of the errors are underestimated, we assume they are all underestimated by the same factor $S$. If we scale up all the input errors by this factor, the $\chi^2$ becomes $K-1$' (ref. \cite{Workman:2022ynf}, p. 16)\end{quote} 

Finally,  given that experiments are grouped  by collider, the PDG approach assumes for each collider $c$ some additional common systematic variance $\Delta^2_{c}$  associated with that collider; this group systematic variance is typically referred to as `correlated error' (\cite{Workman:2022ynf}, page 15) because it reflects variance in individual experimental means arising from factors that affect all experiments in the group. 

 For the set of experiments $C$ from collider $c$ the PDG takes the contribution of this correlated error  $\Delta^2_{c}$ to the variance of each individual experiment $i$ in this group to be  
\begin{equation}
\label{eq:D_i_c}
\Delta^2_{i_c} = \Delta^2_{c} \sum_{j \in C}\frac{\sigma^2_{stat_i} +\sigma^2_{sys_i} }{\sigma^2_{stat_j} +\sigma^2_{sys_j}   }
\end{equation}

This means that the relative correlated error in experiment $i$ arising because that experiment is a member of group $C$ is proportional to the relative within-experiment uncertainty in that experiment.  This procedure is motivated by the fact that, in the fixed-effect model, each experimental mean $\overline{X}_i$ for experiments in group $C$ is an estimate of the same mean $\mu_*$; and so the systematic variance of the (inverse variance weighted) mean  for that group is the inverse-weighted average of the corresponding individual systematic variances.
 This procedure has the advantage that, with the modified systematic errors $\Delta^2_{i_c}$, results for each individual experiment may be treated as independent and averaged in the usual way with other data.  

The PDG uses the same procedure for dealing with correlated error associated with collider types $t$ (e.g. hadron or lepton colliders), so that the total  variance for experiment $i$ using collider $c$ of type $t$ in the PDG approach is
\begin{equation} 
\label{eq:PDG_variance}
\hat{\sigma}_i^2 = \begin{cases}
\ H_c^2\left(\hat{\sigma}_{stat_i}^2+ \hat{\sigma}_{sys_i}^2\right)+ \Delta_{i_c}^2+ \Delta_{i_t}^2;  &H_c >  1 \\ \\
\hat{\sigma}_{stat_i}^2+ \hat{\sigma}_{sys_i}^2+ \Delta_{i_c}^2+ \Delta_{i_t}^2; &H_c \leq  1 \\
\end{cases}
\end{equation}
and the overall aggregated estimate is given by the fixed-effect mean $\hat{\mu}_F$ with variance $\hat{\sigma}^2_F$ (Eqs \ref{eq:fixed_est} and \ref{eq:fixed_var}) using these values $\hat{\sigma}_i^2$.

To illustrate: in recent calculations of the world average $W$-Boson mass \cite{Workman:2022ynf} the PDG selected $10$ experimental estimates for aggregation (out of $25$ estimates listed): $4$ from the Tevatron collider and $2$ from the LHC (both hadron colliders) and $4$ from the LEP (a lepton collider).    Each experiment has individual  statistical and systematic uncertainties $\sigma^2_{stat_i}$ and $\sigma^2_{sys_i}$ with additional correlated systematic uncertainties for some colliders $\Delta_{i_c}$ and collider types $\Delta_{i_t}$.   Of the $10$ experiments being aggregated, $2$ were produced by the CDF group and $2$ by the D0 group; CDF and D0 results were themselves aggregated giving $8$ estimates overall, each from a different research group.  The process of rescaling was carried out separately for each collider $c$, with  both Tevatron and LHC results having $Q_c/(K_c-1) < 1$  but the LEP results having $H_c = Q_c/(K_c-1) = 1.2$.  
The  PDG approach calculates $\sigma_i^2$ for each estimate as in Eq \ref{eq:PDG_variance} and takes the inverse-variance weighted average as in Eq \ref{eq:fixed_est} with variance  as in Eq \ref{eq:fixed_var}; the resulting world average is  $80.377 \pm 0.012$ GeV.

\subsection{Problems }

There are  a number of problems with this approach.   The first  is that the rescaling of experimental variance by $H_c$ is inconsistent with the statistical model of Normally distributed error on which the PDG analysis of variance depends.    As discussed previously, given experimental control of major nuisance factors and the assumption of additive effects for minor nuisance factors, from the Central Limit Theorem in a fixed-effect approach we can reasonably assume that individual measurements in an experiment $i$ will follow a Normal distribution $\mathcal{N}(\mu_*,\sigma^2_{stat_i} + \sigma^2_{sys_i})$.       Given this model of additive error, any `missing' or underestimated error must also be additive:  multiplicative rescaling to account for missing error is inconsistent with this statistical model. 

A second problem  concerns the idea of adjusting or modifying experimental data (in this case multiplying experimental estimates of variance by $H_c$) to enforce consistency with an assumed model (in this case, `returning' the  $Q_c$ statistic to the value of $K_c-1$ required by the fixed-effect model).  Such adjustments to data can give a somewhat misleading impression of experimental results even when the rationale and process of adjustment is clearly explained (as in the PDG analysis);  such adjustments also add some unknown level of uncertainty to the data (in this case uncertainty in the difference between the adjusted and the missing variance of the data).

A third  problem concerns the interpretation of the $Q$ sum and the rescaling variable $H$.   As we saw earlier,  the statistic $Q$ approximately follows the distribution $\chi^2_{K-1}$ under the fixed-effect assumption and so the fixed-effect hypothesis is rejected when $Q$ deviates sufficiently from its expected value of $K-1$ under this distribution.  Rather than taking values of $Q$  larger than $K-1$ as a sign that experimental variances $\hat{\sigma}_i^2$ should be rescaled to better match the assumptions of the fixed-effect model (the interpretation taken in the PDG approach) in fact large values of $Q$ are a sign that  the fixed-effect model should be rejected in favour of an alternative random-effect approach.  

A final problem arises with the use (and enforcement by rescaling) of the fixed-effect model.  A general view in the meta-analytic literature is that  variance in latent means across experiments is to be expected: that `it would be surprising if multiple studies [\ldots] all ended up estimating the same underlying parameter' \cite{higgins2008commentary}.  From this perspective the random-effect model should be taken as a default in aggregating results across multiple experiments, especially since the fixed-effect model is in fact nested within the random-effect model: the fixed-effect model is an instance of the random-effect model with the parameter $\sigma_*^2$ fixed at $0$, and so a random-effect analysis can approach the fixed-effect analysis as a special case.

\section{A random-effect alternative }
 
  To develop a random-effect approach to particle physics parameter estimation we must extend the basic model to deal with systematic variance associated with a given experiment ($\sigma_{sys_i}^2$) and to deal with aggregation across subgroups of experiments with correlated error or common systematic variance $\Delta_{c}^2$.   
   Given some group of $K$ exchangeable experiments (with no subgroups), each with individual sample mean $\overline{X}_i$ and sample variance $\hat{\sigma}_i^2$, and with sample variance between means of $S^2$, we estimate between-experiment variance $\sigma_{*}^2$ via variance-component estimator (Equation \ref{eq:sigma_star}).  If our sample variance $\hat{\sigma}_i^2$ is made up of statistical and systematic components ($\sigma_{i}^2 =\sigma_{stat_i}^2 +\sigma_{sys_i}^2$) then in the random-effect model this systematic variance reflects random factors that impact the mean $\mu_i$ of that experiment (that is, random differences between this experiment and others) and thus represents a component of the  between-experiment variance $\sigma_{*}^2$. 
  Since, unlike estimators based on $Q$, the variance-component estimator allows between-experiment variance to be split into parts in the estimation process,  we can  estimate the additional between-experiment variance (beyond that accounted for by the systematic terms) as
   \[  \sigma_{\epsilon}^2  = \sigma_*^2 - \sum_{i=1}^{K} \sigma_{sys_i}^2 \]
  Given this our random-effect aggregate estimate for the overall mean $\mu_*$ is
\begin{equation}
\label{eq:W_est}
\begin{split}
\hat{\mu}_W &= \hat{\sigma}^2_{W} \sum\limits_{i=1}^K \frac{\overline{X}_i}{ \hat{\sigma}_{stat_i}^2+ \hat{\sigma}_{*}^2  } \\ &=\hat{\sigma}^2_{W} \sum\limits_{i=1}^K \frac{\overline{X}_i}{ \hat{\sigma}_{stat_i}^2+ \hat{\sigma}_{sys_i}^2 + \hat{\sigma}^2_{\epsilon} }
\end{split}
\end{equation} 
where the  variance of this estimate is given by the normalising value
\begin{equation}
\label{eq:W_var}
\hat{\sigma}^2_{W}  = \frac{1}{\sum\limits_{i=1}^K \frac{1}{\hat{\sigma}_{stat_i}^2 + \hat{\sigma}^2_{*} }}
\end{equation}

\subsection{Correlated error}

 Consider a group $c$ containing $K_c$ experiments with some common factor (all run on the same collider $c$, for example) where that common factor has an impact on the mean of each individual experiment in the group (and so represents a `correlated error' for that group).  In the random-effect model we represent this situation by assuming some latent group mean $\mu_{*_c}$ such that  means $\mu_i$ for experiments  in group $c$ are distributed  as
 \[\mu_i \sim \mathcal{N}(\mu_{*_c},\sigma_{*_c}^2) \]
 where group means themselves are distributed around the overall mean as
 \[\mu_{*_c} \sim \mathcal{N}(\mu_{*},\sigma_{*}^2) \]
 Here variance $\sigma_{*_c}^2$ reflects factors that impact the individual experimental means $\mu_{i}$ in  group $C$ and so represents between-experiment (but within-group) variance, while variance $\sigma_{*}^2$ reflects factors that impact the group mean $\mu_{*_c}$ (that is, differences between this group and other groups) and so reflects correlated error in experiments in that group.    In the $W$-Boson case such factors could lead to the underlying group mean $\mu_{*_{LHC}}$ for experiments on the LHC collider being different from the group mean $\mu_{*_{LEP}}$ for  experiments on the LEP collider: such random differences at the collider level will impact the means of all experiments on that collider in the same way.  
 
Given this the overall distribution for the mean of experiment $i$ in group $c$ is
 \begin{equation*}  
 \overline{X}_i \sim \mathcal{N}(\mu_*,\sigma_{i}^2 +\sigma_{*_c}^2 + \sigma_{*}^2)
 \end{equation*}
 As before, if $\sigma_{i}^2 =\sigma_{stat_i}^2 +\sigma_{sys_i}^2$ then $\sigma_{sys_i}^2$ is a component of between-experiment (but within-group) variance, giving
 \[  \sigma_{\epsilon_c}^2  = \sigma_{*_c}^2 - \sum_{i=1}^{K_c} \sigma_{sys_i}^2 \] 
 Similarly,   correlated error $\Delta_c$ is a component of between-group variance giving
  \[  \sigma_{\epsilon}^2  = \sigma_{*}^2 - \sum_{c=1}^{K} \Delta_c^2 \] 
   This gives an equivalent overall distribution for the mean of  experiment $i$ in group $c$ of
\begin{equation}
\label{eq:X_i_group} 
\overline{X}_i \sim \mathcal{N}(\mu_*,\sigma_{stat_i}^2 +\sigma_{sys_i}^2+ \Delta_c^2 +\sigma_{\epsilon_c}^2  +\sigma_{\epsilon}^2)
\end{equation}

 Note that the random-effect model with the variance-component estimator is consistent across all levels of estimation.\footnote{To see this, consider that when aggregating $K$  individual measurements to produce an experimental mean there is no within-measurement variance (each measurement estimates itself exactly) so that $\sigma_i^2=0$ and the only variance is the between-measurement variance $\sigma_*^2$. Estimating this using the variance-component approach as in Eq \ref{eq:sigma_star} we get $\hat{\sigma}_*^2 = S^2$ where $S^2$ is the sample variance of our experiment, and so the inverse-variance weighted mean (Eq \ref{eq:random_est} ) is simply $\overline{X}$ (the unweighted mean) and the variance in that mean, calculated as in Eq \ref{eq:random_var}, is simply $S^2/K$ (the sample variance of the mean).}  This model thus applies equally to estimation of the mean and variance in a single experiment from individual measurements, to estimation of the mean and variance of a group of experiments from their experimental means, to estimation of the mean and variance of a group containing nested subgroups from their subgroup means and variances and so on. Between-experiment (within-group) variance $\sigma_{*_c}^2$ can thus  be estimated using the variance-component estimator applied to the set of means of experiments in group $c$, while between-group variance $\sigma_{*}^2$ can be estimated using the same method applied to the set of group means $\overline{X}_c$.   Given such estimates we have 
\begin{equation}
\label{eq:W_est_group}
\begin{split}
\hat{\mu}_W &= \hat{\sigma}^2_{W} \sum\limits_{i=1}^K \frac{\overline{X}_i}{ \sigma_{stat_i}^2 +\sigma_{sys_i}^2+ \Delta_c^2 +\sigma_{\epsilon_c}^2  +\sigma_{\epsilon}^2  } \\
&= \hat{\sigma}^2_{W} \sum\limits_{i=1}^K \frac{\overline{X}_i}{ \sigma_{stat_i}^2 +\sigma_{*_c}^2+ \sigma_{*}^2  }
\end{split}
\end{equation} 
as our estimate of $\mu_*$  where the  variance of this estimate is given by the normalising value
\begin{equation}
\label{eq:W_var_group}
\hat{\sigma}^2_{W}  = \frac{1}{\sum\limits_{i=1}^K \frac{1}{ \sigma_{stat_i}^2 +\sigma_{*_c}^2+ \sigma_{*}^2   }}
\end{equation}
 
\subsection{Differences in correlated error}

It is worth pointing out two substantial differences between these fixed-effect and random-effect approaches to correlated error in grouped experiments.  The first is mathematical, and is seen when we compare the contribution that correlated error $\Delta_c^2$ associated with a group of experiments makes to the individual variance of those experiments.  In the fixed-effect approach the contribution of this correlated error to each individual experiment's variance is approximately $\Delta_c^2/K_c$  (see Equation \ref{eq:D_i_c}).  This follows necessarily from the fixed-effect assumption, under which $\mu_i=\mu_{*_c}= \mu_*$  holds for all $i$ and so  the individual experimental variances $\sigma_i^2$ must sum to the group mean variance $\sigma_{*_c}^2$.  

 In the random-effect model, by contrast,  with correlated error $\Delta_c^2$ for group $c$, each individual experiment in the group has the same  $\Delta_c^2$ variance component around $\mu_*$ (Equation 
\ref{eq:X_i_group}), because in this model each individual mean $\mu_i$ varies around the group mean $\mu_{*_c}$ and so variance in $\mu_i$ around the overall mean $\mu_*$ is the sum of individual and group variances.  In other words, while the contribution of correlated or group error to a given experiment's variance in the fixed-effect model falls with group size, in the random-effect model this contribution is independent of group size (and so is $K_c$ times larger than in the fixed-effect model).    While this difference may seem surprising, it is a necessary consequence of the fixed-effect versus random-effect distinction, arising because the fixed-effect model assumes no variation in means, while the random-effect model assumes variation in both experiment and group means.

The second difference concerns the grouping of experiments.  In the fixed-effect approach experiments can in principle be grouped in any way, because the fixed-effect statistical model makes no specific assumptions about the distribution of experimental results in groups.   The random-effect model, by contrast, assumes that experimental results in a given group will follow some Normal distribution, and a specific grouping scheme is meaningful only when that assumption holds.  To see this, consider a simple situation where we have $K$ experimental results overall and where we are considering splitting this set of experiments into two subgroups $C_1$ and $C_2$, each containing $K_c$ experiments with correlated error $\Delta_c^2$.  Our statistical model for experimental results in subgroup $C_1$ is then 
 \begin{equation*}  
\overline{X}_i \sim \mathcal{N}(\mu_*,\sigma_{i}^2 +\sigma_{*_1}^2 + \sigma_{*}^2)
\end{equation*}
while that for subgroup $C_2$ is
 \begin{equation*}  
\overline{X}_i \sim \mathcal{N}(\mu_*,\sigma_{i}^2 +\sigma_{*_2}^2 + \sigma_{*}^2)
\end{equation*}
and this division into subgroups $C_1$ and $C_2$ will  have a meaningful impact on our statistical analysis of experimental results only when within-group variances $\sigma_{*_1}^2$ and $\sigma_{*_2}^2$ are different: if these values are the same (if $\sigma_{*_1}^2 = \sigma_{*_2}^2$) then our statistical model for experimental results given this grouping scheme is equivalent to the ungrouped model 
\begin{equation}  
\label{eq:normal_model}
\overline{X}_i \sim \mathcal{N}(\mu_*, \sigma_{*}^2)
\end{equation}
albeit with $\sigma_{*_2}^2$ now representing between-experiment rather than between-group variance.  This means that if we have evidence that the statistical model in Equation \ref{eq:normal_model} holds for a given set of experimental results (that is, if the set of experimental means $\overline{X}_i$ is Normally distributed), then we can conclude that no division of those results into distinct subgroups can be meaningful.  Such evidence can be obtained via a statistical test of Normality such as the Shapiro-Wilk test \cite{razali2011power,shapiro1968comparative} as before; and so a given set of experimental means $\overline{X}_i$ can be meaningfully divided into subgroups only when those means give significant evidence against Normality at some level $\alpha$ in such a test.

\section{The $W$-Boson mass}

We illustrate this approach by using the random-effect model to estimate the $W$-Boson mass. In applying this approach to the set of $10$ experimental estimates selected for the PDG fit \cite{Workman:2022ynf,aaij2022measurement,aaboud2018measurement,aaltonen2012precise,abazov2012measurement,abdallah2008measurement,abbiendi2006measurement,achard2006measurement,schael2006measurement,abazov2002improved,affolder2001measurement}, we first ask whether the model assumption of Normally distributed between-experiment variation is reasonable given this data.  These estimates have an overall unweighted mean of $M=80.386$ GeV with standard deviation of $S= 0.06$ GeV; we can assess the Normality of these estimates by asking whether their distribution is consistent with a Normal distribution $\mathcal{N}(M,S^2)$. One standard assessment is via a Quantile-Quantile plot, which compares the experimental estimates in increasing order against the Median order statistics of the standard Normal distribution.  Figure $1$ shows this plot: the clear agreement between experimental and theoretical order values supports the assumption of Normal between-experiment variance.

\begin{figure}[t!h!] 
	\begin{center}
		\scalebox{0.3}{\includegraphics*[viewport= 0 0  750 650]{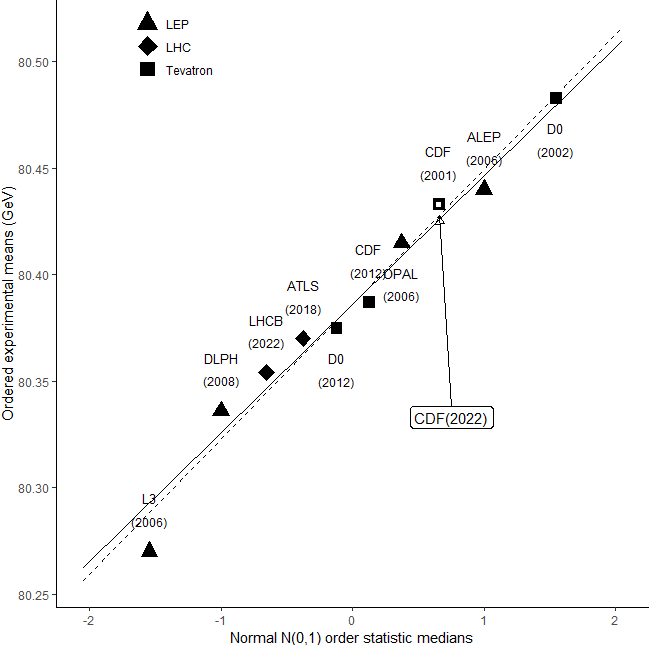}}
	\end{center} 
	\caption{ Quantile-Quantile plot for the $K=10$  $W$-Boson mass estimates used in the PDG fit \cite{Workman:2022ynf}.   Estimates are in increasing order; the $y$ axis gives the estimate, the $x$ axis the corresponding Standard Normal order statistic median (so that for the $k$th lowest mass estimate, the $x$ axis gives the expected median value for the $k$th lowest of $K$ random samples from a Standard Normal distribution).    The  solid line is the line we expect mass estimates to follow if subject to Normal between-experiment error, the dashed line is the linear best fit: the close match between these lines supports the random-effect assumption. The CDF(2022) estimate is shown but not included in the analysis: it is clearly consistent with this assumption of Normally-distributed between-experiment error.  }
\end{figure}

 A second assessment uses the standard Shapiro-Wilk test of Normality; applying the Shapiro-Wilk test to these estimates gives a test statistic of $W=0.983, p= 0.978$ and there is no significant departure from Normality in these estimates.  Given this, and the close agreement with Normal order statistics shown in Figure $1$, we can reasonably conclude that these means are Normally distributed with no meaningful subgroups.  We present two random-effect analyses of this data: the first using the systematic variance values given in the PDG analysis (for illustrative purposes) and the second a full random-effect analysis assuming no such systematic variance values.

 \subsection{Estimate with PDG grouping}

We first calculate an aggregate estimate using the random-effect approach as in Eqs \ref{eq:W_est_group} and \ref{eq:W_var_group} using reported PDG values for $\sigma_{stat_i},\sigma_{sys_i},\Delta_{c_i}$ and $\Delta_{c_t}$, and aggregating  CDF and D0 results giving $8$ estimates overall, each from a different group.  Note that we use the PDG correlated or group error values $\Delta_{c_t}$ and this group aggregation simply to stay as close to the PDG analysis as possible: in fact these  choices are not consistent with the full random-effect model, because the full set of experimental results are consistent with the assumption of Normality and so this grouping is not, in fact, meaningful for these results.   Aggregating these $8$ estimates using the random-effect method gave a world average $W$-Boson mass of $80.376 \pm 0.016$ GeV, with an estimate for additional between-experiment error of $\hat{\sigma}_{\epsilon} =0.024$ GeV.  In comparing the 2022 CDF mass estimate ($80.433 \pm 0.009$ GeV) against the predicted Standard Model value ($80.357 \pm 0.006$ GeV) while taking this additional between-experiment error into account we get a standard deviation for the  difference of $\pm 0.03$ GeV, and the 2022 CDF estimate is approximately $2.6\sigma$ from the SM value (by comparison with the $\sim 7\sigma$ difference obtained under the fixed-effect model).   Similarly comparing the 2022 CDF mass estimate  against our updated world average while again taking this additional between-experiment error into account we get a standard deviation for the  difference of $\pm 0.033$ GeV, and the 2022 CDF estimate is approximately $1.7\sigma$ from the world average.   From this we conclude that the apparent deviation from expectations represented by the 2022 high-precision CDF estimate most likely reflects the fact that the fixed-effect analysis used to assess this deviation ignores an important source of  uncertainty: that arising from between-experiment variation.   

\subsection{Random-effect estimate}

The above $\hat{\sigma}_{\epsilon}$ analysis assumes that all systematic variance estimates ($\hat{\sigma}_{sys_i}^2$, $\Delta_{i_c}^2$ and $\Delta_{i_t}^2$) are accurate and unbiased.  For $W$-Boson mass estimates, however, these systematic variances are dominated by theoretical terms  and so depend on the propagation of variances from other PDG world averages; which by our argument may be subject to some systematic underestimation (because they ignore between-experiment variation).  

 As a check on the above results we can give an alternative random-effect estimate for the $W$-Boson mass, dropping all systematic terms and considering only statistical variance in each experiment (so that $\sigma_{i}^2=\sigma_{stat_i}^2$).  This is an estimate of $\sigma_*^2$ (the total between-experiment variance) rather than  $\sigma_{\epsilon}^2$ (the additional between-experiment variance unaccounted for by the various systematic terms) and so we expect this purely statistical estimate to be consistent with the $\sigma_{\epsilon}^2$ estimate given previously.  Further, this purely statistical estimate avoids various well-known problems with  systematic variance estimation: that such systematics may
 
 \begin{quote}
 	`lack both an unambiguous definition (leading to various recipes to determine these uncertainties) and a clear interpretation (beyond the fact that they are not from a statistical origin) [so that it becomes] a complicated issue to incorporate their effect properly, even in simple situations often encountered in particle physics' (ref. \cite{charles2017modeling}, p 213). 
 \end{quote}  

 Applying the random-effect model to the statistical variances for the $10$ experiments used in the PDG fit (and aggregating the CDF and D0 results as before)  gives a world average $W$-Boson mass of $80.375 \pm 0.016$ GeV with an estimate for total between-experiment error of $\hat{\sigma}_{*} =0.027$ GeV.  Under this analysis the 2022 CDF estimate is  $2.4\sigma$ from the SM value and  $1.6\sigma$ from the updated world average; and so, again, is consistent with both (and in agreement with the $\sigma_{\epsilon}^2$ estimate).

\section{Going beyond the PDG fit} 

The above analyses limit themselves to the $10$  $W$-Boson mass estimates used to produce the PDG fit \cite{aaij2022measurement,aaboud2018measurement,aaltonen2012precise,abazov2012measurement,abdallah2008measurement,abbiendi2006measurement,achard2006measurement,schael2006measurement,abazov2002improved,affolder2001measurement}.  Here we go beyond these results to consider the full set of $25$  $W$-Boson mass estimates listed in the PDG report \cite{Workman:2022ynf}.  We first ask whether this full set of estimates is consistent with the random-effect model assumption of Normality.  Applying the Shapiro-Wilk test for Normality to the means in this set gives a result which conclusively rejects this hypothesis ($W=0.49, p < 10^{-8}$), and so we consider splitting the full set into subgroups.

\begin{figure*}[h!] 
	\begin{center}
		\scalebox{0.35}{\includegraphics*[viewport= 0 0  2050 650]{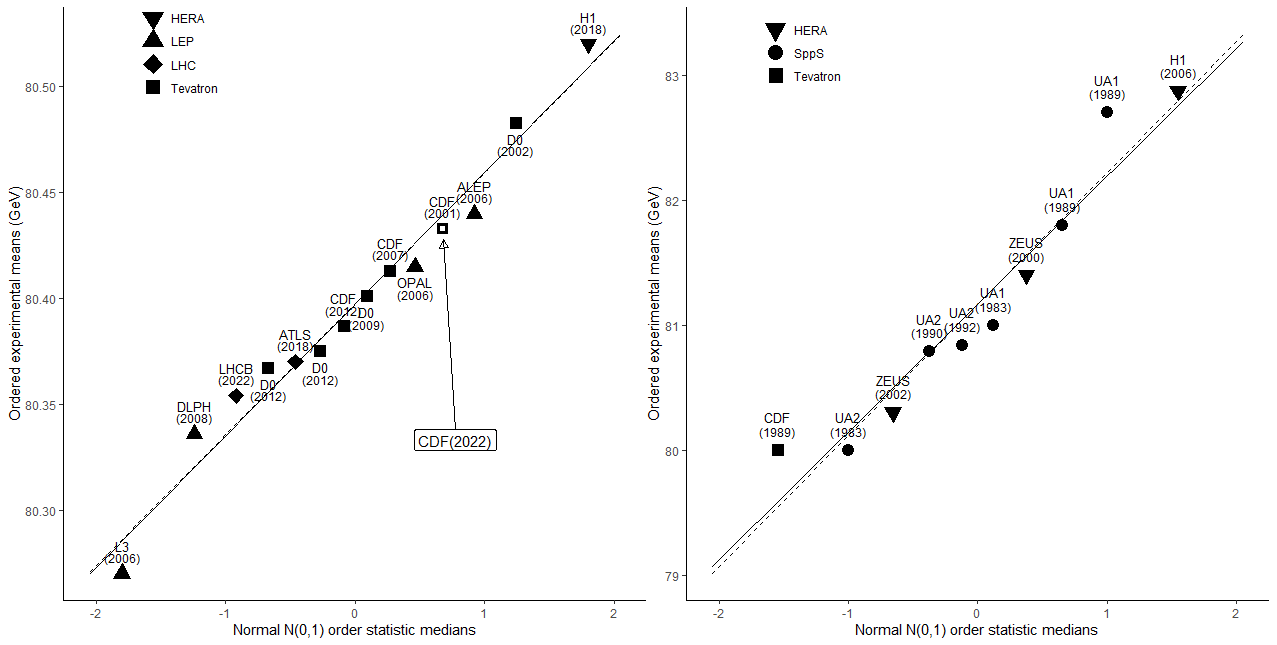}}
	\end{center} 
	\caption{ \textbf{(Left)} Quantile-Quantile plot for the $14$  $W$-Boson mass estimates with $\sigma_{stat_i}^2 \leq 0.12$ in the PDG report \cite{Workman:2022ynf} (all $2001$ or later) and \textbf{(Right)} Quantile-Quantile plot for the $9$  $W$-Boson mass estimates with $ 0.12 < \sigma_{stat_i}^2 \leq 6.7$ in that report (most earlier than $2001$).  Both are consistent with the assumption of Normality, supporting the random-effect approach.  Note the difference in the $y$-axis scale: earlier estimates had much higher within-experiment ($\sigma_{stat_i}^2$) and between-experiment  ($\sigma_{*}^2$) variance than later estimates.     }
\end{figure*}

Our initial presentation described uncertainty in measurement as being dealt with by `controlling for' major nuisance factors and treating minor  nuisance factors as having small and additive effects (so their overall impact on results is described by the Central Limit Theorem). This means that, if we have some set of experiments where major nuisance factors are controlled to some level $x$ (so that all minor factors have some individual impact much less than $x$), then we would expect the results from that set of experiments to be Normally distributed.  Given this, a natural way to place experiments into subgroups is in terms of the level of experimental control, or more specifically in terms of the level of statistical variance or error $\sigma_{stat_i}$: if we have some set of experiments whose level of statistical error is less than some value $x$, we would expect results from those experiments to be normally distributed.

To group experiments based on statistical error we sort the $25$ $W$-Boson mass estimates in the PDG report in increasing order of error $\sigma_{stat_i}$, and find the largest value $x$ such that the set of estimates with $\sigma_{stat_i} \leq x$ are consistent with the assumption of Normal error.  Setting $x = 0.12$ we identify a set of $14$ such estimates, all dating from $2001$ or later,  with a Shapiro-Wilk test on these estimates giving $W=0.98, p= 0.98$, so demonstrating consistency with the Normal assumption (see Figure $2$, Left).   We then continue through the remaining results (again ordered by increasing $\sigma_{stat_i}$ ) to find the largest value $y$ such that estimates with $x <\sigma_{stat_i} \leq y$  are also consistent with Normality.  Setting $y= 6.7$ this generates a second group of $9$ estimates, most dating from before $2001$,  with a test statistic $W=0.91, p= 0.30$, and so also consistent with this assumption.

The first group of $14$ experiments \cite{aaij2022measurement,aaboud2018measurement,aaltonen2012precise,abazov2012measurement,abdallah2008measurement,abbiendi2006measurement,achard2006measurement,schael2006measurement,abazov2002improved,affolder2001measurement,andreev2018determination,abazov2009measurement,aaltonen2007first} had a relatively low degree of statistical variance (mean $\sigma_{stat_i}^2 \approx 0.05$  GeV); the world average $W$-Boson mass estimate produced by applying the random-effect procedure to these results was $80.381 \pm 0.012$ GeV with an estimate for total between-experiment error of $\hat{\sigma}_{*} =0.023$ GeV.  Under this analysis the 2022 CDF estimate is  $2.98\sigma$ from the SM value and  $1.9\sigma$ from the estimated world average, and so, again, is consistent with both.   The second group of $9$ experiments \cite{,aktas2006determination,chekanov2002inclusive,zeus2000measurement,alitti1992improved,ua1990measurement,abe1989measurement,ua11989studies,arnison1983further,banner1983observation} had a much higher degree of statistical variance (mean $\sigma_{stat_i} \approx 3.7$ GeV); the $W$-Boson mass estimate for this group was $80.81 \pm 0.54$ GeV with total between-experiment error of $\hat{\sigma}_{*} \leq 0$ (indicating that uncertainty in individual estimates was too high to give an effective estimate for between-experiment error).  Our overall conclusion, based purely on this statistical analysis and not including any domain-specific information about the characteristics of these experiments, is that the aggregated result from the first group of $14$ more recent high-precision experiments gives a better aggregated estimate for the $W$-Boson mass.   
 
In the random-effect model we would expect that, given a set of experiment with approximately the same level of control over nuisance factors (and so roughly the same degree of  statistical within-experiment variance), means obtained in those experiments would follow a Normal distribution with between-experiment variance reflecting that level of control.   The fact that this set of $25$ results can be cleanly divided into $2$ groups based on within-experiment variance and that means in those two groups are distributed consistently with a Normal distribution thus supports this random-effect approach.  In is interesting to note that these two groupings in some ways demonstrate the development of techniques for estimating $W$-Boson mass, with  statistical error falling over time and with a relatively natural step-change around $2001$, most likely reflecting the very rapid growth in computing infrastructure in high-energy physics towards the end of the $90s$ \cite{boccali2019computing}.

\section{Sources of variance}

It may be useful at this point to compare the fixed-effect and random-effect models in terms of their $W$-Boson mass estimates.   Recall that the fixed-effect PDG model gave a world average $W$-Boson mass of $80.377 \pm 0.012$ GeV  \cite{Workman:2022ynf}, while the random-effect estimate produced by selecting the largest Normally distributed set of high-precision results gave a world average of $80.381 \pm 0.012$ GeV with an estimate for total between-experiment deviation of $\hat{\sigma}_{*} =0.023$ GeV.  The world-average values for the two models are very close and the uncertainty in these values is identical ($\sigma = 0.12$ GeV in both cases, with a difference between these averages of $0.003$ GeV $= 0.33 \sigma$).   The primary difference between these two results, then, lies in the estimate for between-experiment variation in the random-effects model (which is not, of course, part of the fixed-effect approach).  This means that while the two models give almost the same world average $W$-Boson mass (and with the same uncertainty in that value), the two models make differing predictions about future experimental measurements of that mass:  where the fixed-effect model predicts that future results  will be distributed around the world average with a standard deviation of $\pm 12$ MeV 
(and so are expected to be tightly clustered around the world average because this model assumes no between-experiment variance)
the random-effect model predicts that future results  will be distributed around that average  with standard deviation of $\pm (12^2+ 23^2) = \pm 26$ MeV (a significantly larger spread, taking the estimated variation in experiments into account).  

Note that this uncertainty in future experimental results, at $\pm 26$ MeV, is around $3$ times the quoted uncertainty in the recent high-precision CDF result ($\pm 9$ MeV).  Given this three natural questions arise:  first, where does this `missing' uncertainty come from? Second, why was it not identified during the various cross-checks performed at the CDF experiment or others?  Third, why is this missing uncertainty assumed to be the same in each experiment?  

Answers to these questions follow from the conceptual difference between the fixed-effect and random-effect perspectives on experimentation.  In the random-effect perspective, experiments are considered as a unit of analysis and seen as varying randomly in a myriad of minor ways, with small differences in planning, design, technique and measurement between each individual experiment.  These small differences or nuisance factors cannot be individually identified or controlled for in cross-checks (each individual difference having an almost negligible effect).  However, these random factors combine additively to act as the source of significant between-experiment variation in results.  If we were able to identify the specific differences between each experiment and all others, we would be able to estimate the uncertainty caused by these nuisance factors for each individual experiment separately (just as we do when controlling for major between-experiment nuisance factors). Since by assumption there are many of these nuisance factors each with a small and random effect, we cannot carry out such control; however, from the Central Limit Theorem we see that the combined effect of these nuisance factors will be approximately the same for all experiments, even though the individual factors impacting each experiment will vary.

\section{Conclusions}

Questions of uncertainty are fundamental to the estimation of `world average' parameters in particle physics.  Our aim in this paper has been to point out various problems with the fixed-effect approach used to estimate this uncertainty  (an approach that is, to the best of our knowledge,  standard in this area of physics) and to suggest, as an alternative, the random-effect meta-analytic approach that has become commonplace in a range of other fields  \cite{shadish2015meta}.   Taking the $W$-Boson mass as an example, we've shown that  mass estimates vary across experiments in a way that is consistent with a Normal error distribution as assumed in the random-effect model, and that recent experimental results \cite{cdf2022high} that appeared to show statistically significant deviation from theoretical predictions and other experimental results (under  fixed-effect analysis) were in fact not significantly different under a random-effect analysis.  

Our  random-effect world average estimate for the $W$-Boson mass is only marginally different from the PDG fixed-effect estimate but shows a fairly large difference in  uncertainty for the results of future experiments.   This increase in uncertainty is a consequence of the between-experiment variance of $W$-Boson mass estimates (which the fixed-effect uncertainty estimate does not include).   We expect that similar increases in uncertainty will hold for some particle physics parameter estimates in the PDG report (those where individual experimental means differ to some degree) but not for others.   More specifically, if for some particle physics parameter we can reject the random-effect hypothesis of Normal between-experiment variation and our estimate for between-experiment variance is $0$ or negative, then for that parameter we can set $\sigma_*^2=0$ and the mean and standard deviation of the estimate will be just as given in the PDG report.  One such parameter is the mass of the electron, which is known to a very high precision.  For the $8$ electron mass estimates used to produce the PDG world average \cite{Workman:2022ynf},  the assumption of Normal between-experiment variation is rejected at $p=0.00015$ in a Shapiro-Wilk test and the component-variance estimator gives a value of $-1.5 \times 10^{-11}$ (marginally less than $0$), with a standard deviation in that estimate of $2.99 \times 10^{-11}$. These results give an assignment of $\sigma_*^2=0$ in our random-effect approach (no between-experiment variance in these estimates) and so a world average that exactly matches the fixed-effect PDG world average.  

We've illustrated this random-effect approach using the variance-component estimator for between-experiment variance (Eq \ref{eq:sigma_star}) because this estimator is relatively simple to use, not biased by the use of sample estimates, consistent at all levels of analysis, and allows us to `separate out' the systematic variance estimates commonly given in particle physics from the overall  between-experiment variance.  This is not necessarily the best such estimator for use in particle physics settings: researchers have proposed a wide range of estimators for between-experiment variance  (see Ref \cite{veroniki2016methods} for an analysis of $16$ such estimators) and the choice of an appropriate estimator for a given set of experiments depends on the characteristics of those experiments and on the aims of analysis.   Researchers working in experimental particle physics are best placed to make this choice:  we hope this paper will encourage such researchers to consider the random-effect approach when analysing and aggregating experimental results.

 \newcommand{\noop}[1]{}

\backmatter

\bmhead{Acknowledgments}

We would like to thank Martin Gr{\"u}newald for a very helpful discussion on an earlier version of this manuscript.

\end{document}